\documentclass{webofc}

\usepackage[varg]{txfonts}   
\usepackage{hyperref}
\usepackage{url}
\hypersetup{colorlinks=true,citecolor=blue,urlcolor=blue,linkcolor=blue}
\begin{document}
%
\title{Vector Meson Dominance and Vector Meson Production}
%
%

\author{\firstname{Craig D} \lastname{Roberts}\inst{1,2}\fnsep\thanks{\email{cdroberts@nju.edu.cn}.  Supported by
National Natural Science Foundation of China grant no.\ 12135007.
} 
}

\institute{School of Physics, \href{https://ror.org/01rxvg760}{Nanjing University}, Nanjing, Jiangsu 210093, China
\and
Institute for Nonperturbative Physics, \href{https://ror.org/01rxvg760}{Nanjing University}, Nanjing, Jiangsu 210093, China
          }

\abstract{
Against a background supported by the three pillars of emergent mass, this contribution challenges the fidelity of vector meson dominance as an instrument for relating the electromagnetic vector-meson (V) production reaction 
$e + p \to e^\prime + V + p$ to the hadronic process $V + p \to V + p$.  It also describes a viable reaction model for exclusive photoproduction of light and heavy vector mesons from the proton, which exposes the $q\bar q$ content of the photon.  That model's successes reveal that it is premature to link extant $V + p \to V + p$ data with, for instance, in-proton gluon distributions, the QCD trace anomaly, or pentaquark production.  Improved reaction theory and more precise data are required before the validity of such links can objectively be assessed.
}
\maketitle
\section{Introduction}
\label{intro}
The Standard Model of Particle Physics (SM) has one obvious mass-generating mechanism, i.e., the Higgs boson.
Its impacts are critical to the evolution of the Universe.
However, considering the sum of renormalisation group invariant (RGI) light quark masses, only 1\% of the proton mass, $m_p$, can be attributed solely to Higgs boson (HB) couplings into quantum chromodynamics (QCD) \cite[Sec.\,2]{Yu:2026poj}.
Evidently, there is another phenomenon in Nature that is very effective at producing mass.  
This has come to be known as emergent hadron mass (EHM) \cite{Ding:2022ows, Achenbach:2025kfx, Binosi:2026tre}.
Alone, EHM explains 94\% of $m_p$; the remaining 5\% is generated by constructive interference between EHM and HB contributions. 
It is natural to ask: What is EHM and does it have a reductionist explanation; if so, is that explanation contained within the SM?

Understanding EHM is important because the proton mass scale, for which it is responsible, appeared just $\sim 1\mu$s after the birth of our Universe and has subsequently had a determining influence on its evolution.  
In fact, EHM can be identified as the background against which the following key questions for 21$^{\rm st}$ century science can be seen:
How do hadron masses arise; 
how are their spins generated; 
and what role, if any, do gluons play in producing emergent features of strong interactions?
In attempting to answer these questions, the international physics community is engaging with an array of high-luminosity, high-energy facilities \cite{Chen:2020ijn, Anderle:2021wcy, Arrington:2021biu, Aoki:2021cqa, Quintans:2022utc, Ai:2025cpj, Ai:2025xop, Messchendorp:2025men, Accardi:2026slw}.
The quest to explain the source of the mass of (almost all) visible matter is a principal pursuit of modern physics.

\section{A Gluon Mass}
Since the only things that distinguish QCD from quantum electrodynamics (QED) are gluon self-interactions, then these interactions should be the key to understanding EHM, if it lies within the SM. 
Indeed, it was argued some 45 years ago \cite{Cornwall:1981zr} that such interactions are sufficient to transmogrify massless gluon partons, used to define the QCD Lagrangian, into gluon quasiparticles, with novel properties that emerge as the consequence of a momentum-dependent mass function, $m_g(k^2)$, which is nonperturbatively generated by gauge sector dynamics. This mass function vanishes in the ultraviolet, viz.\ on any domain for which perturbation theory is a valid tool.
On the other hand, it is large at infrared momenta, being characterised by a single RGI mass scale, whose value is today known with some precision \cite{Cui:2019dwv}: $\hat m_0 = 0.43(1)\,$GeV, i.e., a value that is roughly half the proton mass!
The gluon mass appears as the consequence of a Schwinger mechanism \cite{Schwinger:1962tn} that is active in QCD \cite{Ferreira:2023fva, Binosi:2026tre}.  

The emergence of $\hat m_0 \approx m_p/2$ is truly \textit{mass from nothing}: an interacting theory, built using massless gluon fields, produces dressed gluons whose propagation is modulated by a mass function that becomes large as the gluon attempts to draw away from the spacetime centre of colour neutrality.  
It is also a QCD fact, being predicted by both continuum and lattice Schwinger function methods.  
The question is: Can the emergence of a gluon mass be empirically verified?  It is anticipated that programmes at modern facilities \cite{Chen:2020ijn, Anderle:2021wcy, Arrington:2021biu, Aoki:2021cqa, Quintans:2022utc, Ai:2025cpj, Ai:2025xop, Messchendorp:2025men, Accardi:2026slw} will deliver the data necessary to supply an answer. 

\begin{figure}[t]
\leftline{\includegraphics[clip, width=0.49\textwidth]{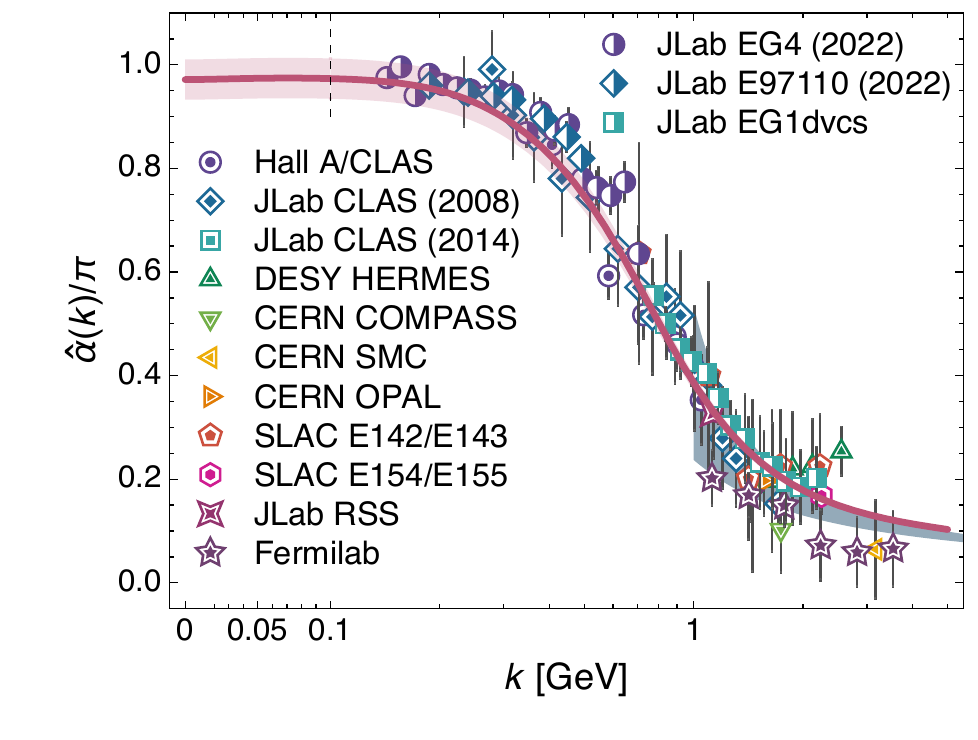}}\vspace*{-31.7ex}

\rightline{\includegraphics[clip, width=0.50\textwidth]{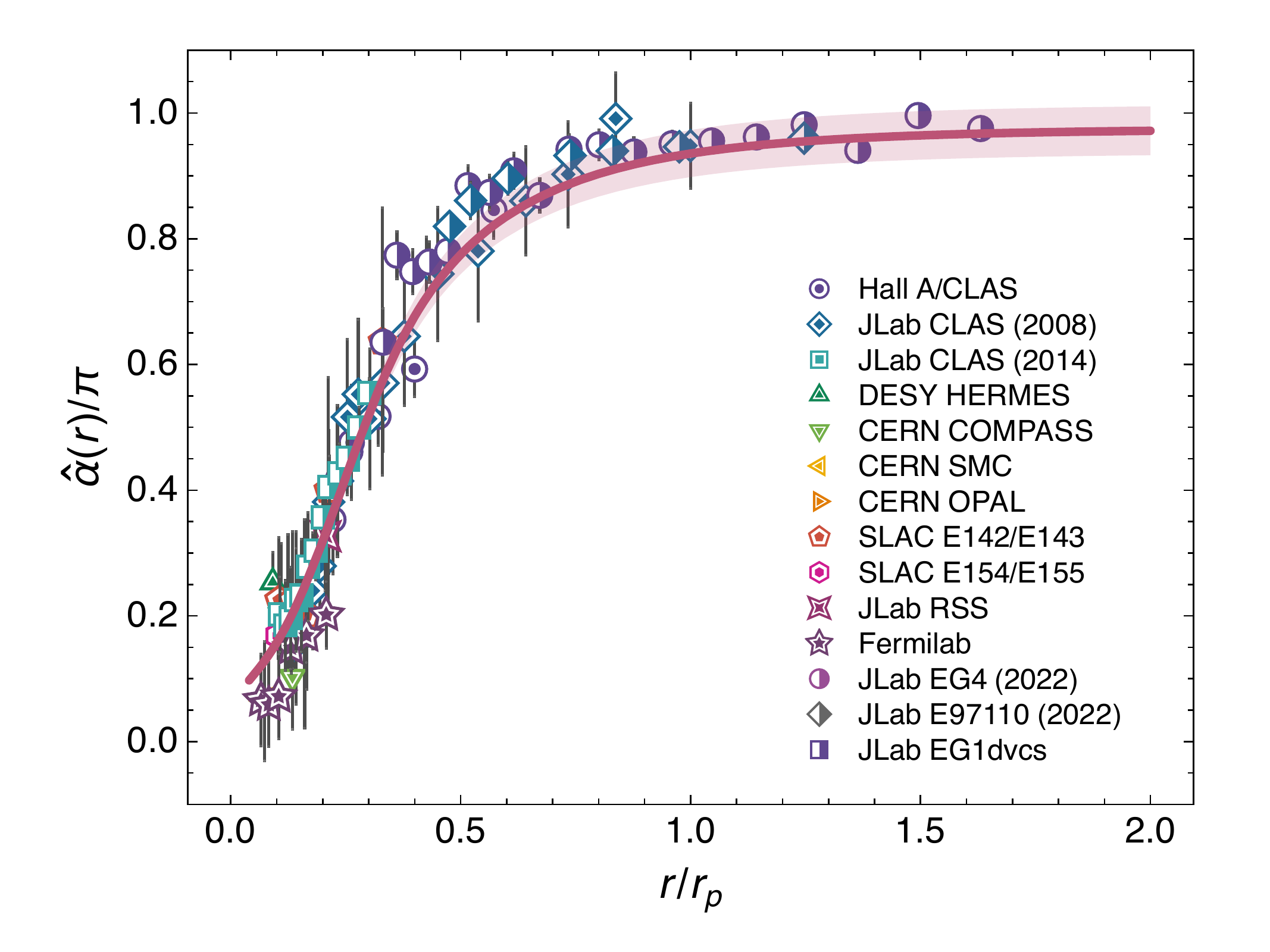}}
\caption{\label{Falpha}
Left Panel.
QCD process-independent (PI) effective charge, $\hat{\alpha}(k)$.
It is calculated by combining the best available results from continuum and lattice analyses of QCD's gauge sector \cite{Cui:2019dwv}.
%
%
Data:  process-dependent charge $\alpha_{g_1}$ \cite{Deur:2022msf, Deur:2023dzc}, defined via the Bjorken sum rule.  
(The sources are listed in Ref.\,\cite[Fig.\,3]{Ding:2022ows}.)
%
Right Panel.  PI charge in Fourier conjugate space, i.e., as a function of spacetime separation, with $r_p$ being the proton electric charge radius.
(Images courtesy of D.\,Binosi.)
}
\end{figure}

\section{Process Independent Effective Charge}
In a non-Abelian quantum field theory, there are infinitely many ways to define a running coupling \cite{Deur:2023dzc}.  
However, following emergence of the gluon mass in QCD, one can proceed to define and calculate the unique analogue of the Gell-Mann -- Low running coupling that is so familiar from QED \cite{GellMann:1951rw}, viz.\ the QCD process-independent (PI) charge, $\hat\alpha(k^2)$. 
The path to this charge is laid upon a foundation provided by the pinch technique and background field method \cite{Binosi:2009qm, Abbott:1980hw}; and it leads \cite{Cui:2019dwv} to the parameter free predictions reproduced in Fig.\,\ref{Falpha}.

This PI charge agrees with that computed in perturbation theory on $k^2 \gtrsim m_p^2$.
On the complementary domain, however, the running slows and is eventually imperceptible, with the coupling reaching a finite far-infrared value \cite{Cui:2019dwv}: $\hat\alpha(k^2=0)/\pi = 0.97(4)$.
These features owe to the existence of $\hat m_0$.  Indeed, it is antiscreening induced by massless gluon partons that causes the QCD coupling to increase with increasing interparticle separation (Fig.\,\ref{Falpha}\,-\,right).  
However, at separations greater than $1/\hat m_0$, the system is populated by quasiparticle gluons, which are massive; so, decouple from quantum loops.  Decoupling eliminates antiscreening, the coupling saturates, and QCD becomes practically conformal in the infrared.

Figure~\ref{Falpha} also compares the predicted PI running coupling with data on the directly measurable process-dependent effective charge defined via the Bjorken sum rule, $\alpha_{g_1}$ \cite[Sec.\,4.3]{Deur:2023dzc}.
On the domain for which QCD perturbation theory is valid, $r/r_p\lesssim 1/4$, the ratio of these two couplings is unity up to corrections with strength $\alpha_{\rm \overline{MS}}(r^2)/20$, where $\alpha_{\rm \overline{MS}}$ is a standard perturbative QCD coupling.
On $r/r_p \gtrsim 1/4$, since both charges are defined via operators that have no overlap with isospin singlet quantities, many dynamical contributions that could distinguish between them are eliminated.
This explains why they continue to follow very similar trajectories with increasing $r$: the Bjorken effective charge also saturates, with $\alpha_{g_1}/\pi = 1$ on  $r\gg r_p$.

The PI charge is a strong candidate for that object which represents the interaction strength in QCD at any given momentum scale.  It is plain from Fig.\,\ref{Falpha} that this charge is finite for all exchange momenta, viz.\ there is no Landau pole.  
The behaviour of $\hat \alpha$ indicates that QCD is a mathematically well-defined, four-dimensional quantum field theory.

\section{Vector Meson Dominance}
\begin{figure}[h]
\centering
\sidecaption
\includegraphics[width=0.35\textwidth, clip]{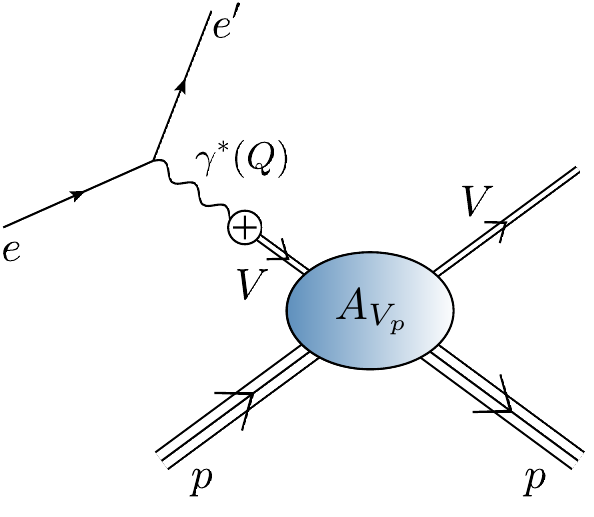}
\caption{Electromagnetic production of a vector-meson from the proton: $e + p \to e^\prime + V + p$.
%
There are three stages.\\
(\textit{a}) $e \to e^\prime + \gamma^\ast(Q^2)$.\\
(\textit{b}) $\gamma^\ast({\rm any}\;Q^2) \to V(Q^2_{\rm new} = {\rm on \; meson \; mass\;shell})$.\\
(\textit{c}) $V({\rm on \; shell}) + p \to V({\rm on \; shell}) + p$.\\
The VMD transition $\gamma^{(\ast)}(Q)\to V$ in (\textit{b}) is indicated by the crossed-circle and is typically assumed to occur with a momentum-independent strength, $\gamma_{\gamma V}$ \cite{sakurai:1969, Fraas:1970vj, Kroll:1967it}.
\label{Fvmd}}       
\end{figure}

The VMD \emph{Ansatz} was introduced before the development of QCD for use in analysing energetic electromagnetic interactions of light hadrons, viz.\ states with masses not much different from $m_p$ \cite{sakurai:1969, Fraas:1970vj, Kroll:1967it}; see Fig.\,\ref{Fvmd}.
It assumes that the photon current is practically indistinguishable from the vector meson current, a notion that is expressed in the field current identity:
\begin{equation}
J_\mu^\gamma = V_\mu \, m_V^2/[2 \gamma_{\gamma V}] ,
\end{equation}
where $m_V$ is the vector meson mass, $\gamma_{\gamma V}$ is a phenomenological coupling strength, and $V_\mu$ is the vector meson field.  It asserts that ``To a very good approximation the entire hadronic electromagnetic current operator is identical with a linear combination of the known neutral vector-meson fields.''
It is still used today and for a much wider range of systems because the alternative is to develop a nonperturbative reaction theory that can explain $q +\bar q$ scattering from hadron targets into vector-meson final-states.  
That remains a challenge. 
Notwithstanding, the lack of a sound alternative (accepted reaction mechanism) does not validate the VMD expedient.
Its fidelity should be reconsidered with QCD constraints in mind.

Commonly, VMD assumes that a photon, which is, at best, real, but is generally spacelike, so that $Q^2 \geq 0$, transmutes into an on-shell vector-meson, with (potentially deeply) timelike momentum $Q^2 = -m_V^2 < 0$; and that the $Q^2 \geq 0$ strength and character of the transition in Fig.\,\ref{Fvmd}(b) is identical with that measured in the real process of vector-meson decay, 
$V \to \gamma^\ast(Q^2 = -m_V^2) \to e^+ e^-$, i.e., the $\gamma_{\gamma V}$ transition coupling is supposed to be fixed at its on-shell value, acquiring no momentum dependence.  This supposition contradicts QCD dynamics.

Consider the photon vacuum polarisation scalar.  The following is a simple fact: 
\begin{equation}
Q^2 \Pi(Q)  \stackrel{Q^2 + m_V^2 \simeq 0}{=} 
\bar e_V^2  \frac{2 f_V^2 m_V^2}{Q^2+m_V^2} \,,
\label{VacPol}
\end{equation}
where $f_V$ is the measured $V\to e^+ e^-$ decay constant.  
Thus, on $Q^2 + m_V^2 \simeq 0$, the timelike photon is indistinguishable from the vector-meson.
This is merely the statement that $e^+ e^-$ collisions, with a tuned centre-of-mass energy, can be used to produce vector mesons.
However, VMD asserts that  the photon is also indistinguishable from the vector-meson on 
$Q^2 \simeq 0$, where, using Eq.\,\eqref{VacPol}, $Q^2 \Pi(Q^2) = 2 \bar{e}_V^2 f_V^2/m_V^2$.
This is a very different statement, which is plainly false because the photon is massless and the associated Ward-Green-Takahashi identity demands $\left. Q^2 \Pi(Q^2)\right|_{Q^2 \simeq 0} \equiv 0$.

There are many theoretical tests to which the VMD assumption can be subjected \cite{Xu:2021mju}: it fails them all.  
Most directly, the following identity is always true:
\begin{align}
 \epsilon^\lambda \cdot \Gamma^\gamma(k;Q) & \stackrel{Q^2+m_V^2\simeq 0}{=}
 \frac{2 f_V m_V }{Q^2+m_V^2} \Gamma_V^\lambda(k;Q) ,
 \label{RatioReal}
\end{align}
where $\Gamma^\gamma(k;Q)$ is the photon-quark vertex, $\epsilon^\lambda$ is the $V$-meson polarisation four-vector, and $\Gamma_V^\lambda(k;Q)$ is the on-shell $V$-meson bound state amplitude.  
VMD asserts that this identity is also true on $Q^2 \simeq 0$.  
However, on $Q^2 \simeq 0$, the photon-quark vertex has three independent terms, whereas 
$\Gamma_V^\lambda(k;Q)$ has eight.
Since $3\neq 8$, VMD cannot be true.  

There is no model-independent (objective) way to estimate and/or correct for the degree by which VMD distorts interpretation of $e + p \to e^\prime + V + p$ reactions in terms of $V+p\to V+p$.  
It is only the latter that can potentially be used to draw connections with, e.g., the in-proton glue distribution. 
Thus, today, for instance, no available data on vector meson photoproduction contains any information about the composition of pentaquark states or gluon parton distributions in the proton.  

 \section{Exclusive photoproduction of $J/\psi$}
The $J/\psi$ contains no valence degrees of freedom in common with the proton.
It has thus long been widely held \cite{Krein:2017usp, Lee:2022ymp} that the dominant mechanisms underlying $J/\psi$ photoproduction must involve some expressions of gluon physics within the target proton and, perhaps, the incipient vector meson, and/or the exchange of (perhaps infinitely many, correlated) gluons between the partonic constituents of each.  (This latter is a Pomeron-like effect.). 
The process is topical because some have argued \cite{Kharzeev:1995ij} that there is a connection between near-thres\-hold $J/\psi$ photoproduction and the in-proton expectation value of the QCD trace anomaly; hence, insights into the character of EHM.
However, any trace-anomaly connection now appears remote \cite{Du:2020bqj, Xu:2021mju, Sun:2021pyw, Tang:2024pky, Sakinah:2024cza}.
Notwithstanding these remarks, there is a continuing desire to interpret $\gamma + p \to V + p$ photoproduction data in terms of gluon physics; hence, elucidation of the underlying reaction mechanism is crucial.

\begin{figure}[h]
\centering
\sidecaption
\includegraphics[width=0.55\textwidth, clip]{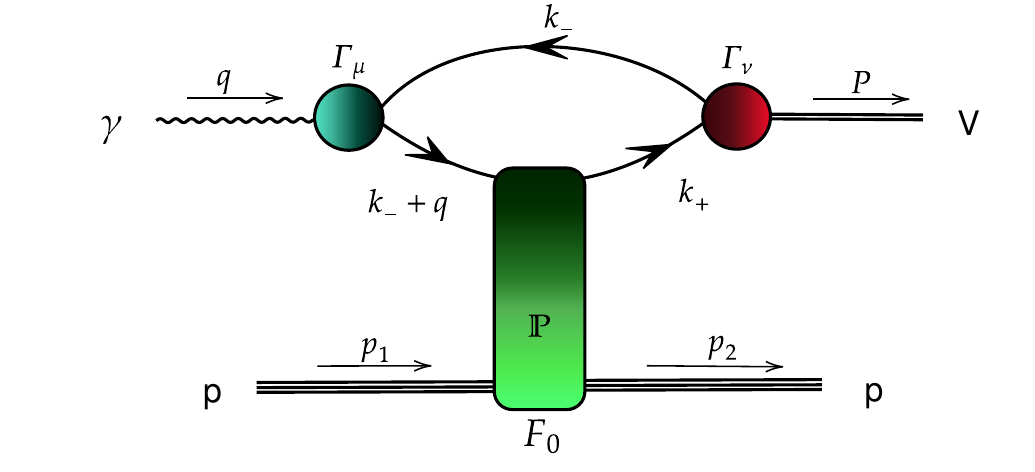}
\caption{
Pomeron-related reaction model for $\gamma + p \to V + p$, referred to hereafter as $\mathbb P$-dyn, in which the $q \bar q$ component of the dressed-photon is probed by Pomeron exchange with the proton, producing an on-shell vector meson, $V$.
Legend. 
Green rectangle describes Pomeron exchange.
Solid black curve -- $S_q$, dressed $q$ quark propagator.
Shaded lighter-blue circle -- $\Gamma_\mu^\gamma$, photon-quark vertex.
Shaded red circle -- $\Gamma_\nu$, $V$-meson Bethe-Salpeter amplitude.
\label{FPom}}       
\end{figure}

As noted above, one cannot reasonably use VMD to interpret $e + p \to e^\prime + V + p$.  
At low-$t$, such reactions are diffractive; hence, a Pomeron-related reaction model can be employed.
In this case, vector meson production proceeds as sketched in Fig.\,\ref{FPom} \cite{Tang:2024pky}:
the incoming photon fluctuates into a (virtual) $q\bar q$ pair;
that pair interacts with the proton via exchange of a Pomeron; 
and the $q \bar q$ pair subsequently coalesces into the final-state, on-shell vector meson.
The analysis is extend to $\rho$, $\phi$, $\Upsilon$ photoproduction reactions, from threshold to very high energies,  in Ref.\,\cite{Tang:2025qqe}.

    Modern continuum Schwinger function methods (CSMs) enable parameter-free prediction of the quark loop structure and associated momentum dependence in Fig.\,\ref{FPom}.  Nine independent scalar functions are produced.  This is called the $\mathbb P$-dyn model.   When the quark loop is replaced by a contact interaction, the $\mathbb P$-am model is obtained. 
Four parameters are associated with $\mathbb P$-exchange in $J/\psi$ production \cite{Tang:2024pky}.
Those defining the Pomeron trajectory, $\alpha_{0,1}$, were fixed twenty years ago by high-$W$ data from HERA.
The Pomeron--light-quark coupling, $\beta_{\ell}$, is fixed by the HERA high-$W$ $\gamma p \to \rho^0 p$ photoproduction differential cross section; and that for the $c$ quark, $\beta_c$, by the HERA  high-$W$ $\gamma p \to J/\psi p$ forward differential cross section.  
Notably, only high-$W$ data are used, i.e., data well away from threshold.  

Other reaction models have been proposed.  
For instance, there is one which attempts to describe $J/\psi$ photoproduction using generalised parton distributions \cite[GPD]{Guo:2023pqw}  and therewith relate the process to in-proton gluon distributions and/or gluon gravitational form factors.
Another \cite[DCC]{Sakinah:2024cza} adopts a Pomeron plus dynamical coupled channels approach, in which final state interactions (FSIs) dominate the near-threshold cross section, so obscuring any possible connection with in-proton gluon distributions and/or gluon gravitational form factors. 

\begin{figure}[h]
\centering
\sidecaption
\includegraphics[width=0.55\textwidth, clip]{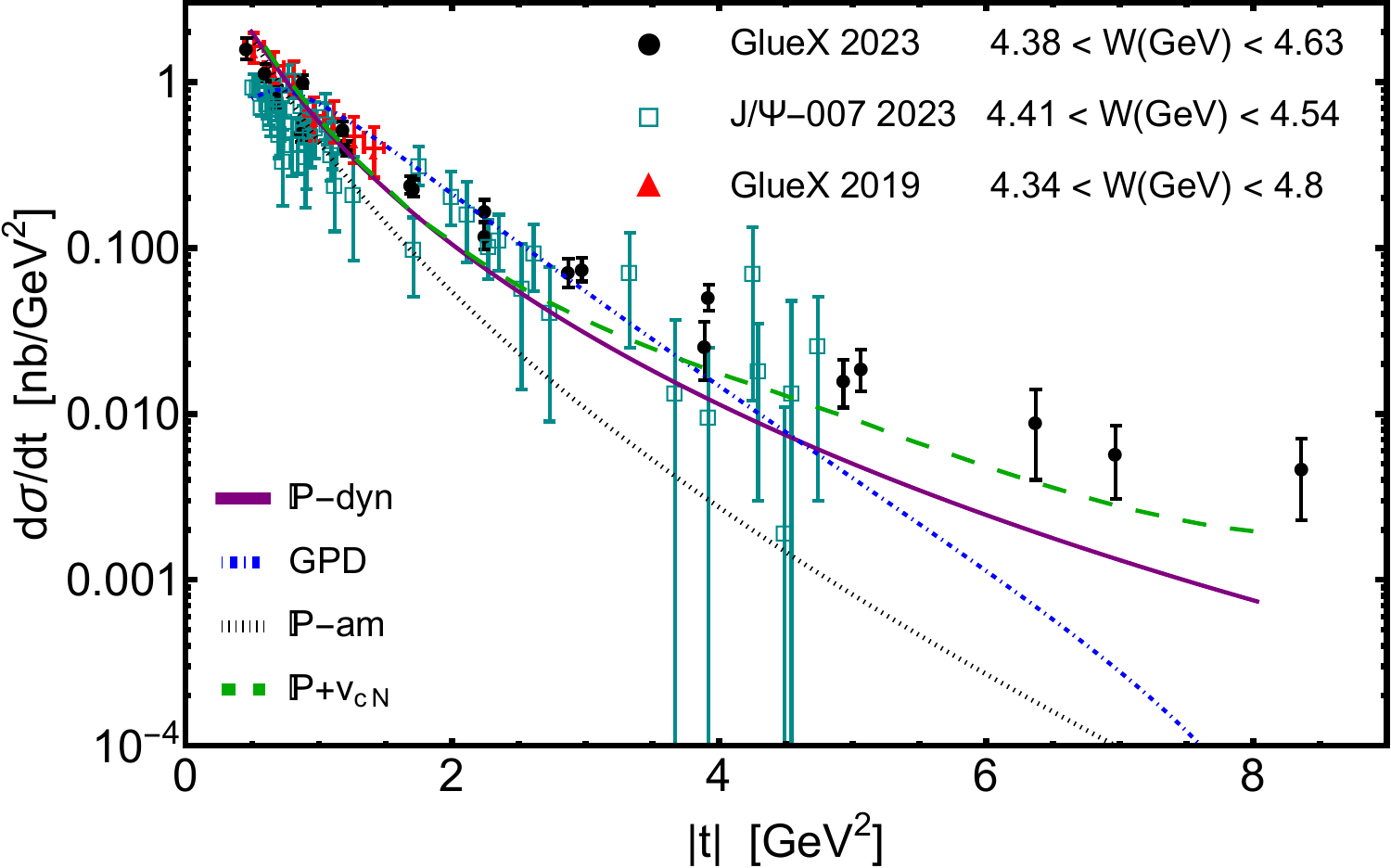}
\caption{
Differential cross-section for $\gamma p \to  J/\psi p$.
Solid purple curve -- $\mathbb{P}-$dyn \cite{Tang:2024pky};
dotted purple curve -- $\mathbb{P}-$am \cite{Tang:2024pky};
dashed green curve --  Pomeron + $J/\psi N$ scattering model \cite{Sakinah:2024cza};
dot-dashed blue curve -- GPD model \cite{Guo:2023pqw}.
%
Data:
$4.43< W/{\rm GeV} < 4.8$ -- \cite[GlueX \,2019]{GlueX:2019mkq};
$4.38< W/{\rm GeV} < 4.63$ -- \cite[GlueX\,2023]{GlueX:2023pev};
$4.41< W/{\rm GeV} < 4.54$ -- \cite[$J/\psi$-007]{Duran:2022xag}.
(All theory curves computed using $W=4.5\,$GeV.)
\label{FigJpsi}}       
\end{figure}

Figure~\ref{FigJpsi} provides a comparison between the four reaction models mentioned above.  
Regarding the GPD model \cite{Guo:2023pqw}, one immediately sees an over-fitting issue: the differential cross section is concave with a maximum at low-$|t|$.  This appears because the modellers had sufficiently many parameters to attempt a reconciliation of \cite[GlueX\,2023]{GlueX:2023pev} and \cite[$J/\psi$-007]{Duran:2022xag} data.  Given the tension between these separate data sets, it is unclear whether such an attempt should be made.
All other models predict a convex differential cross section; hence, remain viable.  
Notably, the $\mathbb P$-am model largely underestimates the data.  This highlights the importance of retaining the quark loop dynamics expressed in Fig.\,\ref{FPom}.  
The $\mathbb P$-dyn model delivers a $\chi^2/$degree-of-freedom$=7$.  By this measure, it is better than the DCC model \cite{Sakinah:2024cza}, for which $\chi^2/$degree-of-freedom$=8$.

\begin{figure}[h]
\centering
\sidecaption
\includegraphics[width=0.55\textwidth, clip]{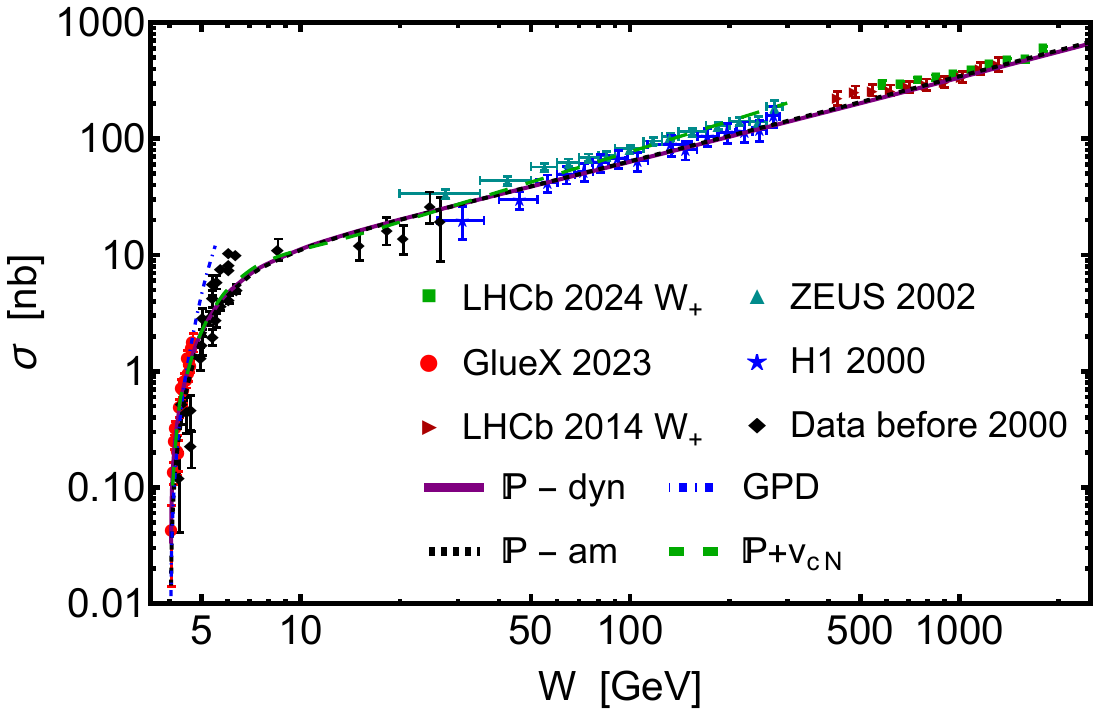}
\caption{
$W$ dependence of total $J/\psi$ photoproduction cross section: at threshold, $W^{J/\psi} = 4.04\,$GeV.
  Solid purple curve -- $\mathbb{P}-$dyn \cite{Tang:2024pky};
  dashed black -- $\mathbb{P}-$am \cite{Tang:2024pky};
  dashed green curve --  Pomeron + $J/\psi N$ scattering model \cite{Sakinah:2024cza};
  dot-dashed blue curve -- GPD model \cite{Guo:2023pqw}.
  Data:
  blue stars -- \cite[H1\,2000]{H1:2000kis};
  cyan up-triangles -- \cite[ZEUS \.2002]{ZEUS:2002wfj};
  red right-triangles -- \cite[LHCb\,2014]{LHCb:2014acg};
  red circles -- \cite[GlueX 2023]{GlueX:2023pev};
  green squares --  \cite[LHCb\,2024]{LHCb:2024pcz};
  black diamonds -- pre-2000 \cite{Camerini:1975cy, Gittelman:1975ix, Shambroom:1982qj, ZEUS:1995kab, E687:1993hlm, H1:1996gwv, Amarian:1999pi}.
\label{FJpsiTotal}}       
\end{figure}

The total cross section for $\gamma p \to  J/\psi p$ is drawn in Fig.\,\ref{FJpsiTotal}.
Both the $\mathbb P$-dyn and DCC models deliver a good description of all available data, with the $\chi^2$ for $\mathbb P$-dyn being somewhat better \cite{Tang:2024pky}.
On the other hand, the GPD model \cite{Guo:2023pqw} does not: in addition to being overfitted for the differential cross section, the GPD model is inapplicable on $W > 5\,$GeV.
Given these observations, it is premature to connect $\gamma p \to J/\psi p$ photoproduction data with anything derived from or connected with GPDs.
The best description of available data is provided by the CSM reaction model sketched in Fig.\,\ref{FPom}.  
Crucially, the $(W,t)$-dependence of the quark loop is important for a good description of the differential cross section near threshold.
The successes of the $\mathbb P$-dyn reaction model suggest that the $q \bar q$ structure of the photon and $J/\psi$ controls the photoproduction reaction dynamics and that available data have no sensitivity to in-proton gluon distributions and/or gluon gravitational form factors. 
The desire to make such connections should be restrained whilst development of reaction models continues and
higher precision data are accumulated.

CSM analyses of proton gravitation form factors exist \cite{Yao:2024ixu, Binosi:2025kpz}.  
They predict the following ordering of proton radii: 
\begin{equation}
r_{\theta}^p = 0.960(36){\rm fm} > r_{\rm charge}^p = 0.887{\rm fm} 
> r_{\rm mass}^p = 0.81(5) r_{\rm charge}^p > r_{\rm mech}^p = 0.72(2){\rm fm} ,
\end{equation}
where $r_{\theta}^p$ is sometimes called the proton scalar radius and $r_{\rm mech}^p$ is the radius of the in-proton normal force distribution.  (Details are available elsewhere \cite[Sec.\,3]{Binosi:2025kpz}.)
There is no objective interpretation of a species decomposition of these radii.
Indeed, such species decompositions may well contain no information beyond that which is already available from the Poincar\'e-invariant gravitational form factors themselves; see, e.g., Refs.\,\cite[Fig.\,5]{Yao:2024ixu}, \cite[Eq.\,(8)]{Binosi:2025kpz}.  The proton $D$-term is predicted to be $D_p(0) = -3.11(1) \approx 3.2 \times D_\pi(0)$, where $D_\pi(0)$ is the in-pion $D$-term \cite{Xu:2023izo}.
These predictions are confirmed by a phenomenological analysis of some available data \cite{Cao:2024zlf}.

\section{Perspective}
The three pillars of EHM are 
the emergence of a mass scale, $\hat m_0$, in the gauge sector of QCD; 
the existence of a process-independent running coupling, which is finite at all length scales owing to the presence of $\hat m_0$; 
and that, acting in concert, these two features lead to the generation of a dressed light-quark mass function, which dynamically breaks chiral symmetry on a scale that is sufficient to explain the proton mass. 
Continuum Schwinger function methods are today identifying the expressions of these phenomena in a large array of measurable phenomena in hadron physics.  

A principal motivation for these efforts is the need to understand if/how the simple QCD Lagrangian can explain the emergence of hadron mass and structure, and, ultimately, atomic nuclei.  
The precise data needed to test any conjectured explanation are expected to become available during operations of modern high-luminosity, high-energy facilities.  
A validated explanation will move science into a new realm by, perhaps, proving QCD to be the first mathematically sound four-dimensional quantum field theory ever contemplated.  In such a case, QCD may form the basis for steps that lead far beyond the Standard Model.

%

\end{document}